\documentclass{aa}
\usepackage[varg]{txfonts}
\usepackage{graphicx}
\usepackage{natbib,twoopt}
\usepackage{textcomp}
\usepackage{xcolor}
\usepackage[breaklinks=true]{hyperref} 

\newcommand{\Mm}{~\mathrm{Mm}}

\title{Structure and dynamics of the internetwork solar chromosphere: results of a small-scale dynamo simulation}
\author{ D. Przybylski\inst{1}, R. Cameron\inst{1}, S.K. Solanki\inst{1}, M. Rempel\inst{2}, S. Danilovic\inst{3}, J. Leenaarts\inst{3}}
\authorrunning{Przybylski et al.}
\begin{document}
\institute{Max-Planck Institute for Solar System Research, 37077 G\"{o}ttingen, Germany
                       \and
            High Altitude Observatory, NSF NCAR, P.O. Box 3000, Boulder, Colorado 80307, USA
             \and
            Institute for Solar Physics, Dept. of Astronomy, Stockholm University, AlbaNova University Centre, SE-10691 Stockholm, Sweden
            \mail{przybylski@mps.mpg.de}
          }
\bibliographystyle{aa}

\abstract{
The heating and structure of the solar chromosphere depends on the underlying magnetic field, among other parameters.
The lowest magnetic flux of the solar atmosphere is found in the quiet Sun internetwork and is thought to be provided by the small-scale dynamo (SSD) process.
}
{
We aim to understand the chromospheric structure and dynamics in a simulation with purely SSD generated magnetic fields.}
{
We perform a 3D radiation-magnetohydrodynamic (rMHD) simulation of the solar atmosphere, including the necessary physics to simulate the solar chromosphere.
No magnetic field is imposed beyond that generated by an SSD process.
We analyse the magnetic field in the chromosphere, and the resulting energy balance.}
{
Plasma at chromospheric temperatures reaches high into the atmosphere, with small, transient regions reaching coronal temperatures.
An average Poynting flux of $5\times10^6~\mathrm{erg\;cm}^{-3}$\;s$^{-1}$ is found at the base of the chromosphere.
The magnetic field in the chromosphere falls off more slowly  with height than predicted by a potential field extrapolation from the radial component of the photospheric field.
Starting in the middle chromosphere, the magnetic energy density is an order of magnitude larger than the kinetic energy density and, in the upper chromosphere, also larger than the thermal energy density.
Nonetheless, even in the high chromosphere, the plasma beta in shock fronts and low-field regions can locally reach values above unity.}
{
The interactions between shocks and the magnetic field are essential to understanding the dynamics of the internetwork chromosphere.
The SSD generated magnetic fields are strong enough to dominate the energy balance in the mid-to-upper chromosphere.
The energy flux into the chromosphere is $8.16\times 10^{6}~\mathrm{erg\;cm^{-2}\;s^{-1}}$, larger than the canonical values required to heat the quiet sun chromosphere and corona.
Possibly due to the limited box size, the simulation is unable to maintain a million-degree corona.
}

\maketitle

\section{Introduction} \label{sec:introduction}

The structure and dynamics of the solar chromosphere is determined by the local magnetic environment. 
The magnetic field couples the solar interior to the atmosphere, controlling the supply of energy and mass from turbulent near-surface convection.
Many mechanisms have been proposed for the heating of the chromosphere, including magnetoacoustic and Alfv\'enic waves, spicules, swirls/tornados, reconnection driven eruptive events and jets.
A large fraction of the solar surface is covered by the internetwork, i.e. the space between the magnetic network in the quiet Sun.
This paper deals with the internetwork chromosphere. 

Observations of the internetwork chromosphere show clear signatures of acoustic shocks, seen as the 'Bright Grains' in CaII H\&K lines \citep{rutten_1991_bright_grains,carlsson_1997_bright_grains}.
Localised brightenings are also seen where small-scale magnetic loops reach into the chromosphere \citep{martinez_gonzalez_2009_smallscaleemergence}.
Internetwork fields cover a large fraction of the solar surface, and provide a significant amount of the magnetic flux at the solar surface \citep{gosic_2016_IN2}. 

At the photosphere, magnetic field measurements utilising Zeeman and Hanle diagnostics reveal ubiquitous small-scale fields \citep{dewijn_2009_smallscalefieldreview}. In radiation-magnetohydrodynamic (rMHD) simulations of the solar convection zone, these quiet-sun fields are generated by a small-scale dynamo (SSD) mechanism \citep{Voegler_2007_dynamo,PietarilaGraham_2010_SSD,Moll_2011_SSD}. 
Simulations including SSD fields produce a close match to magnetic field strengths inferred from observations with the Hinode spectropolarimeter \citep{lites_2011_hinodeSSD,danilovic_2010_MURaMHinode_QS_LD,danilovic_2016_SSDHinode} and Hanle measurements \citep{JTB_2004_hanlefield,dPA_2018_Hanle_QS,zeuner_2024_hanle}.
The temporal-variation of the IN fields is also consistent with an SSD process \citep{buehler_2013_hinodeSSD}.
This dynamo-process is not restricted to the near-surface layers, and is present throughout the convection zone 
\citep{hotta_2015_ssd,hotta_2021_differentialrotation}.
Indeed, recirculation in the deeper convection zone is required to match observationally inferred magnetic field strengths  \citep{rempel_2014_numerical}.
 The large Reynolds and magnetic Reynolds number of the sun put simulation with realistic diffusivities beyond current computational capabilities. \citet{brandenburg_2019_reverseddynamo} compared direct numerical simulations (DNSs), with explicit viscous and resistive diffusivities, to large-eddy simulations (LESs) with numerical diffusive terms. They demonstrated similar behaviour of the SSD between the DNSs and LESs at different Reynolds and magnetic Reynolds numbers. \citet{warnecke_2023_SSD} showed that in DNSs approaching the solar parameter regime the small scale dynamo is able to operate. For a detailed review of the small scale dynamo in the context of quiet Sun magnetism, see \citet{rempel_2023_SSDreview}.

Previous simulations of the quiet sun are often performed with a weak imposed field.
\citet{martinez_sykora_2019_chromo_field} have performed a simulation of the quiet-sun at high resolution, using the Bifrost code \citep{Gudiksen_2011_Bifrost}, including also a chromosphere and corona.
This simulation included a weak ($2.5~\mathrm{G}$) guide field, which is amplified by near-surface convection. 
This simulation includes non-LTE radiative transfer effects, however it does not include a non-equilibrium (NE) treatment of hydrogen ionisation. 
They find a chromosphere dominated by shocks, which are able to amplify the magnetic field.
In a subsequent paper, \citet{martinez_sykora_2023_IN_NE_GOHM} find that including both NE ionisation and a generalised Ohm's law leads to a hotter chromosphere.

A number of studies have investigated the capability of SSD generated fields to supply an energy flux into the solar atmosphere, sufficient to counteract chromospheric radiative losses.
\citet{amari_2015_SSDchromoheating} demonstrated that the SSD model could provide a high enough Poynting-flux to heat the solar chromosphere. 
They use a Boussinesq SSD simulation of the near-surface convection zone, coupled to a simplified atmosphere model. 
They found the observationally inferred magnetic field strengths could provide a Poynting-flux of $4\times 10^6~\mathrm{erg\;cm^{-2}\;s^{-1}}$ into the upper atmosphere, and the re-connection of these fields could reproduce chromospheric features.
Similar results were found in a LTE MURaM simulation of the quiet sun in \citet{tilipman_2023_photospheric_PF}.
They found that the Poynting flux in the lower atmosphere is significantly higher than the canonical values required to heat the chromosphere $2.28\times 10^7~\mathrm{erg\;cm^{-2}\;s^{-2}}$.

At the photosphere, the gas pressure dominates and the effect of SSD fields on the convective dynamics is relatively minor. 
Simulations with SSD fields have slightly smaller granules, reduced convective upflow velocities and form bright points \citep{bhatia_2023_SSD2}.
In the chromosphere, as the plasma beta drops below unity, the presence of the magnetic field comes to dominate the dynamics.
Simulations of the corona and chromosphere including a magnetic field self-consistently generated through a SSD mechanism have been performed with the MURaM code \citep{rempel_2017_extension, chen_2021_campfires,chen_2022_dopplershifts}.
These simulations are large enough to contain a couple of super-granules and self-consistently generate network-scale fields.
The resulting simulation contains a self-consistent corona.
These simulations are performed at relatively low resolutions ($50-120~\mathrm{km}$) and use an Equation of State (EoS) with a local thermodynamic equilibrium (LTE) treatment of hydrogen ionisation.
Despite the success of photospheric SSD simulations in reproducing photospheric observations, the ability of these fields to heat the chromosphere has not yet been self-consistently modelled. 

In this paper we investigate the structure and dynamics of the magnetic field in the internetwork chromosphere utilising a NE rMHD simulation.
We study the dynamics and structure of the internetwork chromosphere generated purely by an SSD.
There is no net flux imposed on the simulation, nor any prescribed flux emergence.
This work includes for the first time a SSD simulation including an NLTE treatment of the radiative losses and hydrogen ionisation in the chromosphere.
The simulation has a high resolution and low diffusion in order to resolve the small scale features of the quiet sun.
We study the impact of SSD fields in the chromosphere, and the structure of the resulting chromosphere and corona of the model.

In Section \ref{sec:numerics} we explain the numerical methods and experimental setup. 
In Sect. \ref{subsec:simulation_overview} we provide an overview of the simulation, including the resulting photospheric field.
The chromospheric magnetic field generated by the SSD mechanism is shown in Sect. \ref{subsec:chromofield}, the energy balance in Sect. \ref{subsec:energybalance}, and the resulting Poynting flux in Sect \ref{subsec:poynting_flux}.
Finally, in Sect. \ref{sec:discussion} we discuss the results and in Sect. \ref{sec:conclusion} present the conclusions.

\section{Numerical approach}\label{sec:numerics}

We perform a 3D radiative MHD simulation using the MURaM code \citep{voegler_2005_muramcode}. 
The code has been extended to include the physics required to model the corona \citep{rempel_2017_extension} and the non-local thermodynamic equilibrium (NLTE) chromosphere \citep{2022_przybylski_chromoMURaM}.
We use the hyperbolic heat conduction and an Alfv\'en speed limiter that incorporates a semi-relativistic treatment with a reduced speed of light \citep{Boris_1970_BC}, implemented through a projection operator in the momentum equation \citep{gombosi_2002_semirelativistic_MHD}.
In order to accurately treat the solar chromosphere we  include a 3D scattering multigroup scheme \citep{skartlien_2000_multigroup,hayek_2010_radiative}, chromospheric line losses \citep{carlsson_2012_approximations}, and a time-dependent treatment of hydrogen ionisation \citep{sollum_thesis_1999,leenaarts_2007_nonequilibrium}.
The simulation code solves the MHD equations in a cartesian geometry for density ($\rho$), velocity ($\mathbf{v}$), magnetic field ($\mathrm{B}$), and hydrodynamic energy ($E_{\mathrm{hydro}} = E_{\mathrm{int}}+\rho \frac{v_x^2+v_y^2+v_z^2}{2}$), where $E_{\mathrm{int}}$ is the internal energy.
We do not solve for the total energy to prevent numerical errors when plasma beta becomes very low in the corona.
The simulations do not include a generalised Ohm's law \citep{braginskii_1965_PI}. 
We provide a summary of the numerical setup in Appendix \ref{app:simulation}.

For the EoS we follow the method described in \citet{2022_przybylski_chromoMURaM}.
We use a pre-tabulated LTE EoS in the interior, merging it to the NE EoS at a pressure of $p_{\mathrm{cutoff}}=3\times 10^{5} ~\mathrm{dyn\;cm}^{-2}$.
The NE EoS includes a time dependent treatment of hydrogen, and $H_2$ molecules, while all other elements are treated in LTE.
We solve the advection of the hydrogen populations with the bulk fluid motions, the system of rates governing the ionisation/recombination and excitation/de-excitation of atomic hydrogen, and the gas-phase transitions forming/destroying molecular hydrogen.
A Newton-Raphson method is used to solve the system of rate equations for the number density $n$ for a 5 level hydrogen atom, protons and molecular hydrogen, as well as temperature ($T$) and electron number ($n_e$). 
The pressure can then be calculated $p=n k_{\mathrm{B}} T$, where $n$ is the total number density.

We use the FreeEoS package \citep{Irwin_2012_freeeos} to generate the LTE EoS, with the abundances of \citet{asplund_2009_abu}.
For the NE EoS, a slight modification is made to the convergence scheme described in \citet{2022_przybylski_chromoMURaM}.
Due to the low magnetic field strengths in the corona the time-step can reach values of $\Delta t \geq 10^{-2}~\mathrm{s}$.
These large time-steps can cause slow convergence of the NE EoS.
To address this we apply the solution of the hydrogen ionisation in two steps.
First we solve the system of rates with a time-step of $10^{-4}~\mathrm{s}$, which settles the fast rates. 
A second step is then made using the time-step of the MHD solver ($\Delta t_{\mathrm{mhd}}$).
The MHD timestep is determined by the maximum fast-wave velocity in the simulation $\Delta t_{\mathrm{mhd}} = \mathrm{min}\left(\frac{\Delta x}{v_A + c_s + |\mathbf{v}|}\right)$, in terms of the Alfv\'en velocity ($v_A$), the sound speed  ($c_s$), and the magnitude of the fluid velocity ($|\mathbf{v}|$). 
In order to prevent unrealistic temperatures a floor on the temperature is set at $2200~\mathrm{K}$, and on internal energy at $1.6\times 10^{12}~\mathrm{erg\;g^{-1}}$. The heating term does not contribute significantly towards the energy balance of the chromosphere, and acts only to improve the convergence of the NE-EoS at low temperatures.

For the scattering multi-group radiation scheme we use opacities calculated using the MPS-ATLAS package \citep{witzke_2021_mpsatlas}.
The multi-group scheme includes 4 bands with the binning determined by the formation height relative to a reference wavelength.
We use the continuum at $500\;~\mathrm{nm}\;$  as a reference, with bin boundaries located at $\log_{10} \tau_{500}= -2.5,-1.5,-0.5$.

The diffusion scheme follows \citet{rempel_2017_extension} with slight modifications.
This scheme uses a slope-limiter, acting on the primitive variables, combined with a $h$-value (see \citep{rempel_2014_numerical}) which limits diffusion in sufficiently smooth areas.
A higher h-value allows a larger gradient to be reached before limiting occurs, and therefore will give a lower numerical diffusion.
In the photosphere we follow \citep{rempel_2014_numerical} and use a value of 2 for the mass ($h_m$), velocity ($h_v$), energy ($h_E$), and magnetic field diffusion ($h_B$).
For the chromospheric regions, below a density of $5.0\times10^{-9}~\mathrm{g\;cm}^{-3}$ we increase the diffusion (reduce the h-parameter) to $h_m = 1.5$, $h_v=1.0$, $h_E=1.0$ and $h_B=1.0$ to minimise the overshoot at shock fronts.
The reduced mass diffusion, relative to the energy diffusion, prevents the formation of an instability,  which occurs in cold chromospheric shock expansions with strong flows, as described in \citet{2022_przybylski_chromoMURaM}.
This method supersedes the use of increased diffusion in regions with low-$\gamma$, where $\gamma$ is the adiabatic index, which was used in the previous work.
Finally, when the density is lower than $10^{-12}~\mathrm{g\;cm}^{-3}$ we use a value of $h_m=1.5$, $h_v=1.0$, $h_E=1.0$, and $h_B=4.0$ to achieve a high-Prandtl number regime, as described in \citet{rempel_2017_extension}.
We apply diffusion separately to the hydrogen populations and the instantaneous internal energy terms, which includes the non-hydrogen components and the microscopic kinetic energy.
Additional hyper-diffusion in the z-direction is applied to the vertical momentum, density, internal energy and vertical magnetic field. 

The lower boundary condition is the Open Boundary, Symmetric Field (OSb) of \citet{rempel_2014_numerical}.
This boundary condition is open to flows.
The entropy in the boundary cells is interpolated from a stellar structure model of the sun, and the pressure is extrapolated into the ghost cells.
The magnetic field is passively advected (set symmetrically in the ghost cells). 
This allows horizontal field to be advected into the domain, mimicking the effects of recirculation deeper in the convection zone.
The upper boundary is open to outflows, and imposes a potential magnetic field \citep{Rempel_2009_sunspot}. 
The horizontal boundary conditions are periodic.

The simulation is initialised with a horizontally isotropic energy and density profile, based on a G2V stellar model, extending to $1~\mathrm{Mm}$ above the photosphere.
The governing parameters of the simulation are summarised in Appendix \ref{app:simulation}.
A small random velocity field is added and the simulation is run until convection stabilises. 
A zero net flux seed field is added, consisting of a vertical ($B_z$) component, which is set to a random value for each column and is constant in the vertical direction.
This gives a zero net-flux magnetic field with an RMS field strength of $\approx 10^{-3}~\mathrm{G}$.
The simulation was run for two days of solar time to allow the field to saturate.
The behaviour of the SSD is dependent on properties of the turbulent convection, such as the Mach number, Reynolds number, magnetic Reynolds number, and magnetic Prandtl number.
Due to the numerical diffusion scheme used in large eddy simulations, like those performed in this work, it is not possible to unambiguously determine the Reynolds number (based on the electrical conductivity of plasma).
Estimates can be made based on the integral scale of turbulence, and the Taylor microscale \citep{PietarilaGraham_2010_SSD,Moll_2011_SSD}.
These methods typically underestimate the magnetic Reynolds number, giving a value lower than the threshold for the small-scale dynamo instability.
Comparison between large eddy simulations and simulations including an explicit resistivity show that the SSD shows similar behaviour \citep{Voegler_2007_dynamo,brandenburg_2019_reverseddynamo}.

The simulation was then extended to $11~\mathrm{Mm}$ above the surface using a constant atmosphere in hydrostatic equilibrium with an arctan temperature distribution and potential magnetic field extrapolation. The simulation was allowed to evolve for another day with an Alfv\'en speed limit of $100~\mathrm{km\;s}^{-1}$.
Finally the Alfv\'en limiter was relaxed to $3000~\mathrm{km\;s}^{-1}$ and the simulation was run including the NE treatment of hydrogen for 60 minutes.
The simulation was then run for an additional 30~minutes where we saved the state of the atmosphere every 10~s.
The final simulation spans a region of $12\times12\times18~\mathrm{Mm}$, ranging from $-6.6~\mathrm{Mm}$ below to $11.4~\mathrm{Mm}$ above the solar surface.
The simulation includes $512\times 512 \times 900$ grid-points giving a horizontal resolution of $23.4375~\mathrm{km}$ and a vertical resolution of $20~\mathrm{km}$. 

\section{Results} \label{sec:results}

\subsection{Simulation overview}\label{subsec:simulation_overview}

\begin{figure*}[htp]
\begin{center}
\includegraphics[width=18cm]{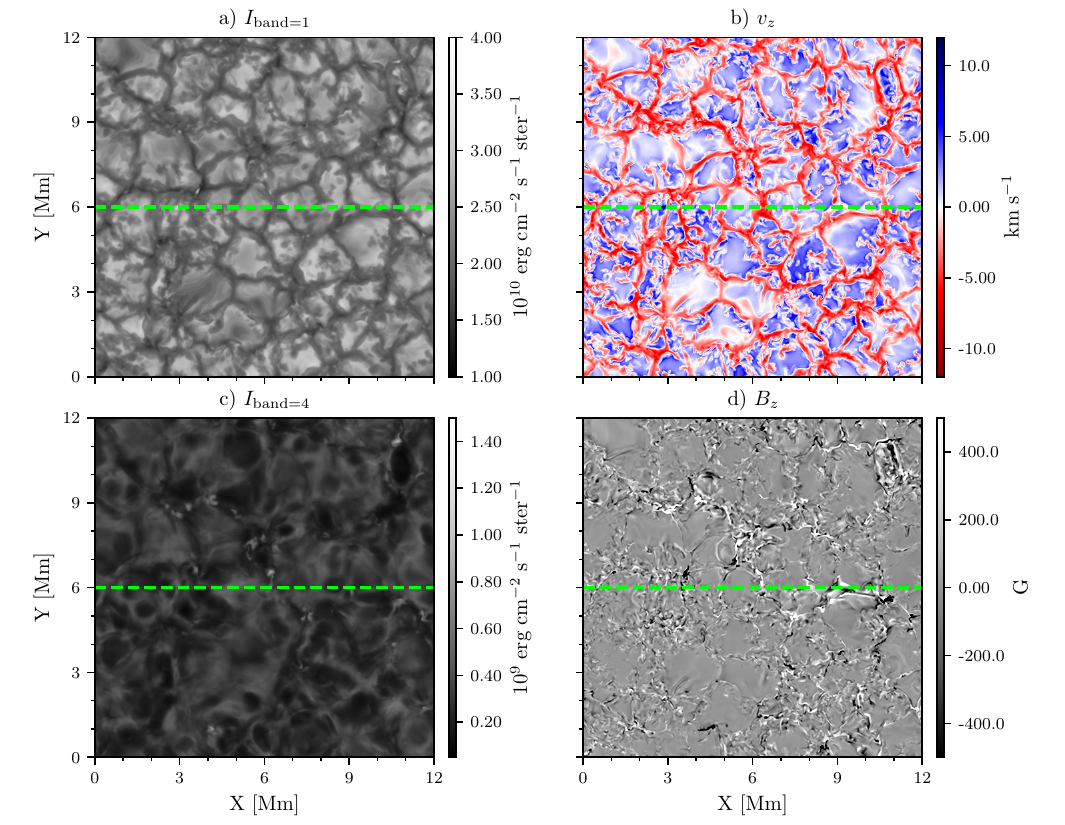}
\caption{A snapshot of the simulation, panel a) The intensity in the continuum (1st) band of the multigroup scheme, panel b) the vertical velocity at $\tau_{500} = 1$, panel c) the intensity in the low-chromospheric (4th) band of the multi-group scheme, and panel d) the vertical magnetic field strength at $\tau_{500} = 1$. The maximum field strength has been saturated to better illustrate the salt-and-pepper field concentrations. The dashed line represents the slice shown in Fig. \ref{fig:slice_atmosphere}\& \ref{fig:Eslice}. \href{https://datashare.mpcdf.mpg.de/s/cdBaEXnLoWpl0o9}{Animation available online}, the image represents $t=0$ of the animation.}
\label{fig:slice_tau1}
\end{center}
\end{figure*}

We first investigate the photosphere and low chromosphere of the simulation in Figure \ref{fig:slice_tau1}.
A typical quiet-sun granulation pattern is seen in the continuum radiation (Panel a) and vertical velocity (Panel b).
At the photosphere, the magnetic field consists of numerous salt-and-pepper concentrations (Panel d).
The vertical velocity and vertical magnetic field are taken on the surface where $\tau_{500} = 1$.
A number of flux concentrations over $1000~\mathrm{G}$ are seen, and the average maximum unsigned vertical field strength over the time series is $|B_z|_{\mathrm{max}} = 2147 \pm 199 ~\mathrm{G}$.
The mean of the horizontally averaged unsigned vertical field is $\left<|B_z|\right>_{\mathrm{havg}} = 71.6 \pm 2.98~\mathrm{G}$ and the RMS magnetic field strength  is $B_{\mathrm{rms}} = 141.3 \pm 7.42~\mathrm{G}$.
The horizontal component of the field at the photosphere, $B_h = \left<\sqrt{B_x^2 + B_y^2}\right>_{\mathrm{havg}} = 103.6 \pm 9.07~\mathrm{G}$.
This is consistent with the inferred field strengths of Hanle measurements of around $130-170 ~\mathrm{G}$ \citep{JTB_2004_hanlefield,zeuner_2024_hanle}, and previous simulations with the MURaM code \citep{rempel_2014_numerical,bhatia_2022_SSD}.
In order to see the upper photosphere we plot in panel c the intensity of the 4th radiation band of the 3D multi-group scheme.
This band consists of frequencies ($\nu$) where the lines become optically thick ($\tau_{\nu}=1$) at heights in the atmosphere with $\tau_{500} < 10^{-2.5}$, including upper-photospheric and chromospheric spectral lines. 
The image reveals a shock interference pattern with persistent bright points above the strong magnetic field concentrations. 

\begin{figure*}[htp]
\centering
\includegraphics[width=18cm]{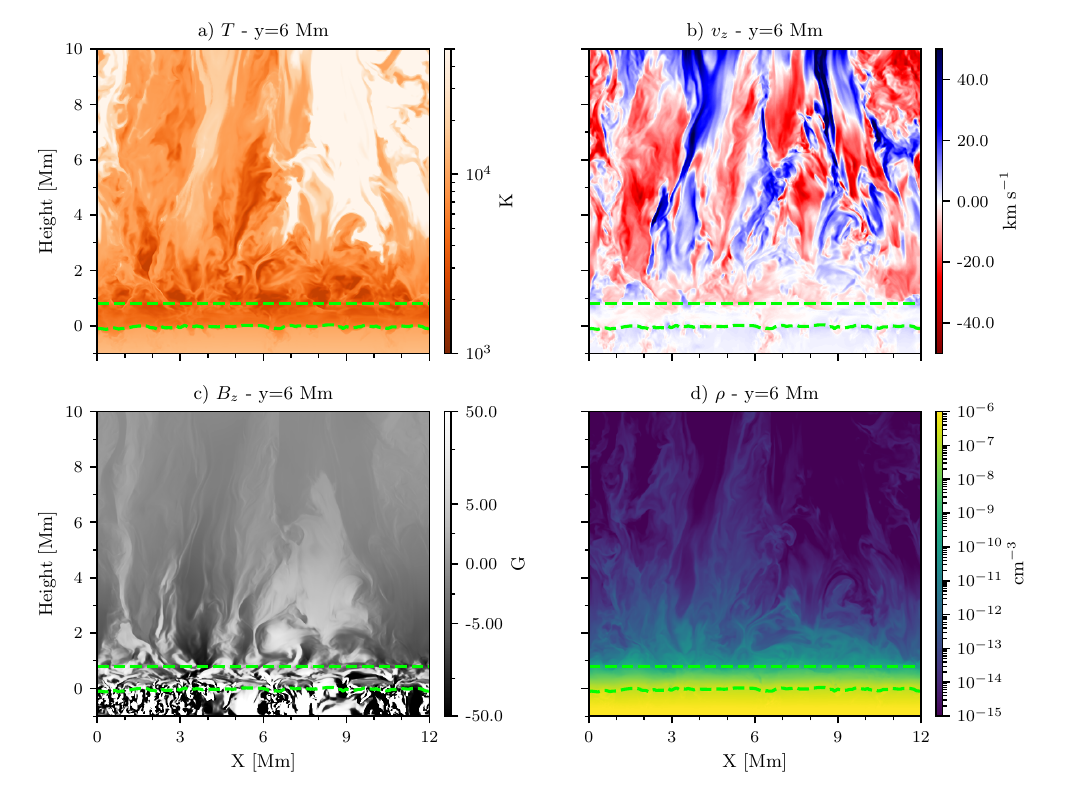}
\caption{A vertical slice through the simulation taken in the xz plane at $y=6~\mathrm{Mm}$.
The figure includes; panel a) Temperature, b) vertical velocity and c) vertical magnetic field, and d) the plasma density.
We use an arcsinh norm for the magnetic field, and limit all panels to highlight the chromosphere.
The horizontal dashed lines represent the photosphere $\tau_{500}=0$ and base of the chromosphere $z=0.8~\mathrm{Mm}$.
\href{https://datashare.mpcdf.mpg.de/s/flU0hJhAJUmPKlh}{Animation Available online}, the image represents $t=0$ of the animation.}
\label{fig:slice_atmosphere}
\end{figure*}

The physical properties of the chromosphere are illustrated in Fig. \ref{fig:slice_atmosphere}.
The chromosphere is seen to be highly structured with large variations in  temperature (panel b) and density (panel d) on small horizontal scales.
This variability makes it difficult to meaningfully separate the chromosphere into regions based on a geometric height above the photosphere.
We instead follow an approach similar to that of \citet{Withbroe_1977_chromoheating} to distinguish between the different layers of the solar atmosphere.
The near-photosphere extends upwards from the surface where $\left<\tau_{500}\right> = 1$, to the temperature minimum.
Following \citet{Withbroe_1977_chromoheating} we define the base of the chromosphere at a height of $\tau_{500}=10^{-5}$.
To simplify analysis, we take a fixed horizontal slice at the height where the horizontally averaged $\left<\tau_{500}\right>_{\mathrm{havg}}=10^{-5}$ as the base of the chromosphere. This slice lies approximately $800~\mathrm{km}$ above the average $\tau_{500}=1$ layer.

Above the temperature minimum, in the low chromosphere, hydrogen is still largely neutral, plasma $\beta$ is predominantly larger than unity, and acoustic shocks dominate the dynamics.
In the mid-chromosphere the ionisation fraction of hydrogen begins to rise, and the plasma temperature increases to $\approx8~\mathrm{kK}$.
In the upper chromosphere, hydrogen is almost fully ionised and the material is heated to near-transition region temperatures, $10~\mathrm{kK}$ to $50~\mathrm{kK}$, as helium is ionised.
A large amount of the upper chromospheric material exists at approximately $10~\mathrm{and}~30~\mathrm{kK}$, the preferred temperatures of the first and second ionisation stage of helium in LTE \citep{golding_2016_NEhelium}.
Finally, above $\approx 50~\mathrm{kK}$ a transition region and corona forms. 

The tenuous transition region and corona that forms in the simulation is strikingly different from previous simulations which include additional imposed fields \citep{carlsson_2016_public,martinez_sykora_2019_chromo_field,breu_2022_loop}. 
When large scale magnetic fields are present, the plasma-$\beta$ is always below zero in the upper atmosphere and the dynamics become field aligned, giving the chromosphere its observed spicular nature. Additionally, in smaller simulations a hot plate is often used to force the formation of a transition region and corona \citep[e.g.][]{leenaarts_2007_nonequilibrium,hansteen_2006_dynamicfibril,rempel_2021_efficient_nonideal}.
The atmosphere in the present simulation is more similar in appearance to the flux-emergence experiment of \citet{Hansteen_2023_MgIIinBF}, or in large quiet sun simulations including network fields \citep{chen_2021_campfires,rempel_2017_extension}.
In these models, chromospheric material can be seen to reach high into the atmosphere.
The structure of the internetwork in these models, in regions with only a small net flux, has not been studied in detail.
Observational evidence exists for a more corrugated transition region. \citet{trujilo_bueno_2018ApJ...866L..15T} state that a more corrugated transition region than present in the Bifrost model is required to match observations from the Chromospheric Lyman-Alpha SpectroPolarimeter (CLASP) rocket experiment \citep{kano_clasp_2017ApJ...839L..10K}.
A study of the \ion{Mg}{ii} h\&k lines in a MURaM Enhanced network model \citep{ondratschek_2024_MgIIEN} came to a similar conclusion.

The animation  of Fig. \ref{fig:slice_atmosphere} reveals a highly dynamic chromosphere.
Large regions of the upper atmosphere exist at transition region temperatures and above.
The bulk of the low-to-mid chromosphere has velocities in the range of $5-10~\mathrm{km\;s^{-1}}$. 
In the upper chromosphere features reaching above $50 ~\mathrm{km\;s^{-1}}$ are seen.
Shock fronts propel chromospheric plasma of $\approx 10~\mathrm{kK}$ into the upper atmosphere, and strong shock expansions frequently reach the minimum allowed temperature of $2200~\mathrm{K}$.
The animation of Fig. \ref{fig:slice_atmosphere} reveals regions of localised heating at the interaction of shocks.
Additionally small scale reconnection events and bi-directional jets can be seen, for example at $x=5~\mathrm{Mm}$, $z=5.5-6~\mathrm{Mm}$, $t=14-15~\mathrm{s}$.
The shocks appear to carry chromospheric material, and magnetic field, into the upper atmosphere. 
The vertical component of the magnetic field (panel c) reveals both persistent loops (for example at $x=10~\mathrm{Mm}$) and highly turbulent, quickly evolving small-scale field structures.
Strong up-flows can be seen to constantly supply magnetic field high into the chromosphere.
For example, at about $x=3~\mathrm{Mm}$, $t=4~\mathrm{minutes}$ a flux emergence event occurs, causing a cool loop to rise up to about $z=5-6~\mathrm{Mm}$.

\begin{figure}[htp]
\centering
\includegraphics[width=8.8cm]{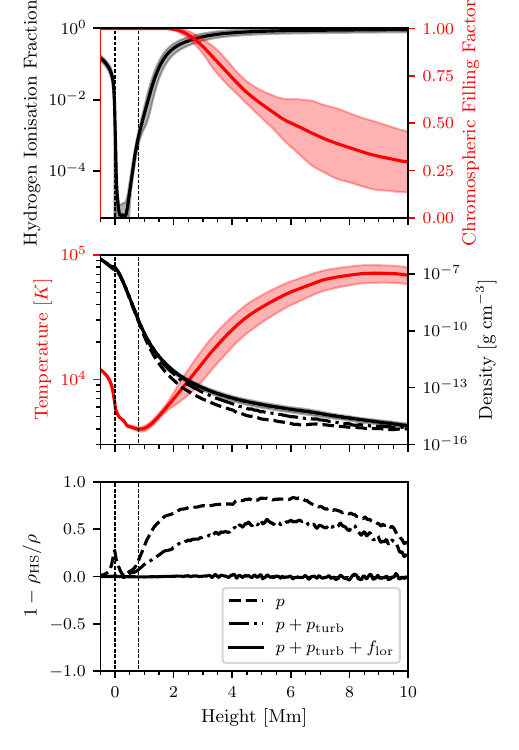}
\caption{Horizontally and temporally averaged thermodynamic properties of the simulated chromosphere. 
Top: the hydrogen ionisation fraction and chromospheric filling factor, which is the area fraction at each height with a temperature below 20kK. Middle: the temperature (red) and density (black). The shaded areas cover two standard-deviations around the mean.
Additionally the density which is recovered from hydrostatic equilibrium is included in the middle panel.
This is calculated including only the gas pressure (dash-dotted), the gas and turbulent pressures (dash-dotted), and the gas pressure, turbulent pressure and magnetic term (indistinguishable from the averaged density).
Bottom: the error of the density calculated recovered with the hydrostatic equilibrium approximation.
The vertical dashed lines represent the photosphere ($z=0~\mathrm{Mm}$) and the base of the chromosphere ($z=0.8~\mathrm{Mm}$).}
\label{fig:avg_chromo_thermodynamics}
\end{figure}

The averaged thermodynamic properties of the chromosphere are plotted in Fig. \ref{fig:avg_chromo_thermodynamics}.
The horizontally averaged hydrogen ionisation fraction (top panel) drops sharply at the photosphere, reaching a minimum value around $z=400~\mathrm{km}$.
The temperature minimum in the simulation is at about $z=800~\mathrm{km}$.
This corresponds closely to the height we have defined as the base of the chromosphere, based on the horizontally averaged $\left<\tau_{500}\right>_{\mathrm{havg}}=10^{-5}$.
The top of the chromosphere is difficult to define due to the large variations of temperature and the presence of chromospheric material high in the atmosphere.
We define a chromospheric filling factor as the area fraction at each height with a temperature below $20~\mathrm{kK}$.
The chromospheric filling factor begins to fall at $2~\mathrm{Mm}$, reaching $50\%$ at $5.5~\mathrm{Mm}$ and falling to $30\%$ at the top of the simulation domain.
The average temperature reaches typical low transition region values, around $20~\mathrm{kK}$ at $3.5~\mathrm{Mm}$.
The average temperature plateaus at about $60~\mathrm{kK}$ towards the top of the simulation domain.
In the upper atmosphere, approximately 10 pixels in a million reach temperatures greater than $1~\mathrm{MK}$.
The spread of temperature increases in the upper chromosphere, as the non-equilibrium ionisation of hydrogen prevents recombination, leading to larger temperature variations.

We investigate the extent to which the averaged chromosphere can be described in pressure balance.
In hydrostatic equilibrium, the density $\rho=-\frac{\partial_z p}{g}$ in terms of the pressure $p$, and gravity $g$.
For a discussion of the hydrostatic balance at the photosphere see \citet{bhatia_2022_SSD}.
In the middle panel of Fig. \ref{fig:avg_chromo_thermodynamics} we compare the averaged simulation density to that calculated including only the gas pressure, to that calculated including the turbulent pressure $p_{\mathrm{turb}} = \rho v_z^2$ and magnetic term $f_{\mathrm{lor}}=\frac{1}{8 \pi}\left(B_h^2 - B_z^2\right)$.
This magnetic term is not the magnetic pressure, but results from the horizontal average of the vertical component of the Lorentz force.
In the chromosphere, the density can only be recovered when including both the gas and magnetic pressure terms.
In the bottom panel we show the error in the density recovered.
When only the gas pressure is included this error is $50~\mathrm{to}~70\%$ in the chromosphere.
Including the turbulent pressure but not the Lorentz force term, reduces the error to between $20~\mathrm{to}~50\%$.
This illustrates that the gas, kinetic and magnetic terms are all important in the structuring of the quiet sun chromosphere.

\subsection{The chromospheric magnetic field} \label{subsec:chromofield}

\begin{figure}[htp]
\centering
\includegraphics[width=8.8cm]{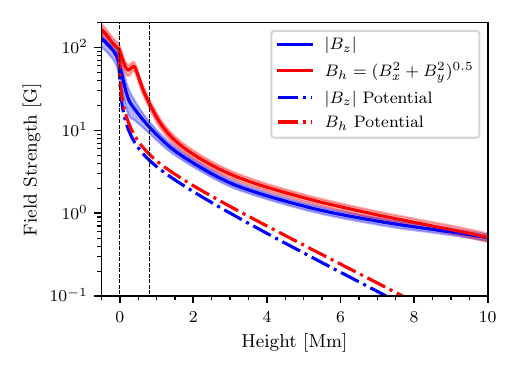}
\caption{Horizontally and temporally averaged values of unsigned vertical field $|B_z|$ and the horizontal field $B_h = \left( B_x^2 + B_y^2 \right)^{0.5}$ through the atmosphere. The shaded regions cover two standard deviations. The dash-dotted lines illustrate the magnetic energy assuming a potential field extrapolation from z=0.}
\label{fig:magnetic_field_avg}
\end{figure}

The simulation shows a number of small kilo-Gauss magnetic field concentrations at the photosphere.
Above the photosphere the magnetic field strength falls off quickly.
By the base of the chromosphere, the variation of the horizontally averaged, horizontal field strength is $\left<B_h\right>_{\mathrm{havg}}= 22.14 \pm 1.34~\mathrm{G}$.
This is a factor of two higher than the unsigned vertical field, $\left<|B_z|\right>_{\mathrm{havg}}=11.24 \pm 0.81~\mathrm{G}$.
The dependence of magnetic field on geometric height in the atmosphere of the simulation is shown in Fig. \ref{fig:magnetic_field_avg}.
We plot the horizontally and temporally averaged values of the unsigned vertical field strength $\left<|B_z|\right>_{\mathrm{avg}}$ and the horizontal field strength $\left<B_h\right>_{\mathrm{avg}}$.
The averaged horizontal field is larger than the averaged unsigned vertical field throughout the atmosphere.
This difference is greatest in the upper photosphere and low chromosphere, which may be a signature of the presence of low-lying loops, as discussed by \citet{schuessler_2008_BhinSSD}.
The horizontal field falls to within a factor of two of the vertical field by $2~\mathrm{Mm}$.
The fact that the two averaged field components approach the same magnitude towards the top of the box could be an effect of the upper boundary conditions.
There is a distinct peak in the $\left<B_h\right>_{\mathrm{havg}}$ about $450~\mathrm{km}$ above the photosphere \citep{rempel_2014_numerical}.
This peak occurs at the minimum of the horizontally averaged RMS velocity above the photosphere.
\citet{rempel_2023_SSDreview} interpret this peak as the height to which turbulent convective motions are able to carry the horizontal magnetic field.
Comparison to a potential field extrapolated from $B_z$ at the photosphere (dash-dotted lines) reveals that the field strength in the dynamic simulation is approximately two times higher in the low chromosphere.
In the upper atmosphere of the simulation the decrease of the magnetic field strength flattens, so that the field is approximately an order of magnitude higher than the potential field by 6 Mm in height.

\begin{figure}[htp]
\centering
\includegraphics[width=8.8cm]{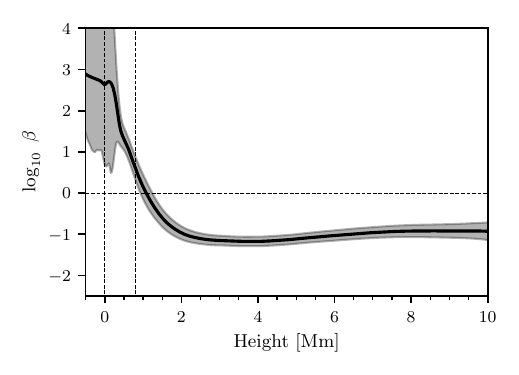}
\caption{Horizontally and temporally averaged values of $\log_{10}\beta$. The shaded region covers two standard deviations. The vertical dashed lines represent the photosphere $(z=0)$ and the base of the chromosphere $(z=800~\mathrm{km})$.}
\label{fig:beta_avg}
\end{figure}

The influence of the magnetic field on the plasma dynamics can be quantified by the plasma-beta $\beta=\frac{8 \pi p}{B_x^2+B_y^2+B_z^2}$.
A high-$\beta$ plasma is dominated by the gas pressure $(\log_{10} \beta > 0)$ and a low-$\beta$ plasma has a larger magnetic pressure.
Horizontally averaged $\log_{10} \beta$ is shown in Fig. \ref{fig:beta_avg} and the shaded region represents one standard deviation. 
At the photosphere, the plasma-$\beta$ is high, values of $\log_{10}\beta$ around or below zero are found only in the inter-granular flux concentrations.
At approximately $1.1~\mathrm{Mm}$ the average $\log_{10} \beta$ becomes negative, and by $2~\mathrm{Mm}$ almost all the plasma is low-$\beta$.
In the mid-to-upper chromosphere the plasma-$\beta$  flattens, as the majority of closed loops turn over and only large scale loops remain.
The limited simulation domain prevents large-scale flux imbalance from occurring, resulting in low field strengths and regions of high-$\beta$ towards the upper boundary of the simulation.
The minimum in the average $\log_{10} \beta$ occurs at about $3-4~\mathrm{Mm}$ above the photosphere, and increases slightly to a value of approximately $\log_{10} \beta=-1$ for the bulk of the atmosphere.

\subsection{The chromospheric energy balance} \label{subsec:energybalance}

\begin{figure}[htp]
\centering
\includegraphics[width=8.8cm]{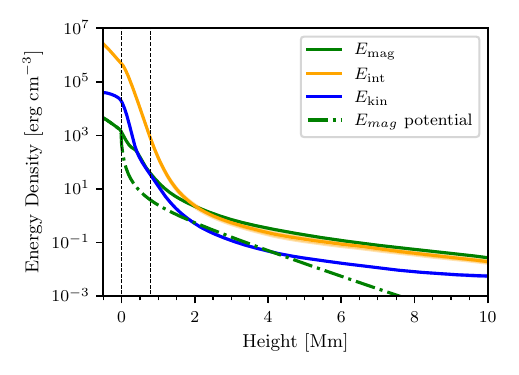}
\caption{Horizontally and temporally averaged values of the kinetic energy (blue), internal energy (yellow) and magnetic energy (green) in the solar atmosphere. The shaded regions cover two standard deviations. We include the magnetic energy of a potential field above $z=0$ (dot-dashed line). The vertical dashed lines represent the height of the photosphere $(z=0)$ and the base of the chromosphere $(z=800~\mathrm{km})$.}
\label{fig:energy_avg}
\end{figure}

The horizontally and temporally averaged energy densities are calculated for the kinetic $E_{\mathrm{kin}} = \rho \mathbf{v}^2/2$, magnetic $E_{\mathrm{mag}} = \mathbf{B}^2/\left(8 \pi \right)$, and internal $E_{\mathrm{int}}$ components, and plotted in Fig. \ref{fig:energy_avg}.
At the solar surface $\left<E_{\mathrm{int}}\right>_{\mathrm{avg}}$ is an order of magnitude higher than $\left<E_{\mathrm{kin}}\right>_{\mathrm{avg}}$, and two orders of magnitude higher than the $\left<E_{\mathrm{mag}}\right>_{\mathrm{avg}}$.
As the density drops the magnetic energy becomes larger than the kinetic energy in the low-chromosphere ($\approx 1~\mathrm{Mm}$), and larger than the internal energy in the mid chromosphere ($\approx 2~\mathrm{Mm}$).
A larger scatter is seen in the internal energy in the chromosphere, as frequent low-temperature post-shock expansions are present in the domain, as well as plasma at chromospheric and coronal temperatures.
The scatter at low energies would be larger without the artificial limits on temperature and internal energy.

In the low chromosphere ($0.5-1~\mathrm{Mm}$) the kinetic and magnetic energies are roughly in equipartition. The kinetic energy drops sharply in the mid chromosphere, where it is typically around an order of magnitude lower than the magnetic and internal energies.
The internal and kinetic energies decrease roughly three orders of magnitude between the photosphere and base of the chromosphere (vertical dashed lines in Fig. \ref{fig:energy_avg}).
This sharp decrease is largely caused by the drop in density above the photosphere, which also falls approximately three orders of magnitude (See Fig. \ref{fig:avg_chromo_thermodynamics}).
Comparison to a potential field, extrapolated from the photosphere (dash-dotted lines) shows magnetic energies from the simulation are around an order of magnitude higher in the low-to-mid chromosphere.
The energy of the potential field remains significantly lower than the kinetic energy aside from a region from $2~\mathrm{Mm}$ to $4.5~\mathrm{Mm}$.
The magnetic energy in the potential field decreases significantly by the upper chromosphere, and is approximately two orders of magnitude lower at $8~\mathrm{Mm}$ above the photosphere.
In contrast, the magnetic energy in the MHD simulation decreases more slowly, in rough equipartition with the thermal energy.

\begin{figure}[htp]
\centering
\includegraphics[width=8.8cm]{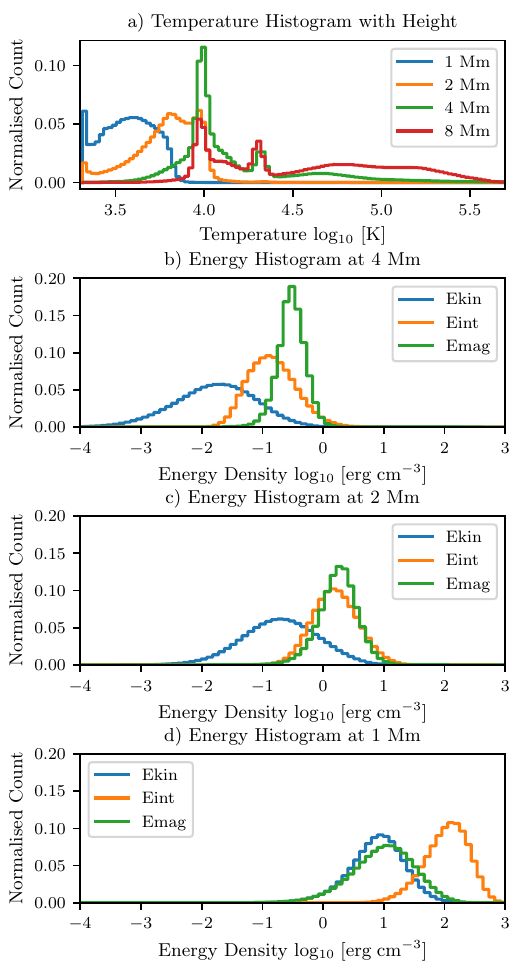}
\caption{Histograms of the energy densities at three different heights in the atmosphere and temperature at four different heights. Panel a) shows a normalised histogram of temperature in the chromosphere at $1~\mathrm{Mm}$, $2~\mathrm{Mm}$, $4~\mathrm{Mm}$ and $8~\mathrm{Mm}$. The lower panels show normalised histograms of internal, magnetic and kinetic energy in the chromosphere, at heights of b) $4~\mathrm{Mm}$, c) $2~\mathrm{Mm}$ and d) $1~\mathrm{Mm}$.}
\label{fig:T_energy_histograms}
\end{figure}

In panel a) of Fig. \ref{fig:T_energy_histograms} we show histograms of the temperature at three different heights.
The histograms show a number of peaks, which shift to higher temperatures at higher heights.
These peaks occur at the temperature where dissociation of molecular hydrogen, or the ionisation of hydrogen and helium occur.
In LTE the peaks are much sharper than in Fig. \ref{fig:T_energy_histograms}, as the transitions are instantaneous, while in NE the slow ionisation/recombination rates in the chromosphere lead to a broader spread of temperatures \citep{Carlsson_2002_dynamichydrogen,golding_2016_NEhelium}.
The peak at $2200~\mathrm{K}$ at $1\mathrm{Mm}$ and at $2\mathrm{Mm}$ is a consequence of the artificial heating term.
At a height of $2~\mathrm{Mm}$ two peaks are seen, at the preferred ionisation temperature of hydrogen, and the first ionisation temperature of helium.
At $4~\mathrm{Mm}$, there is a third peak at the preferred temperature of \ion{He}{ii} ionisation \citep{golding_2016_NEhelium}.
Large regions of the upper atmosphere, at $8~\mathrm{Mm}$, have temperatures of above $100~\mathrm{kK}$.
However a persistent million degree corona is unable to form, with only approximately 1\% of the plasma above the photosphere reaching a million kelvin.

To understand the distribution of energy with height, we plot in Fig. \ref{fig:T_energy_histograms} histograms of the different energy components at heights of $1~\mathrm{Mm}$, $2~\mathrm{Mm}$, and $4~\mathrm{Mm}$.
Above $4~\mathrm{Mm}$ the different energy components decrease slowly with height, with the relative difference remaining similar.
The internal energy histogram keeps a similar distribution at different heights in the atmosphere. 
The magnetic energy histogram becomes narrower with height, while the kinetic energy histogram broadens with height.
The narrowing of the magnetic energy histogram is expected as the magnetic field becomes space filling and plasma-$\beta$ is smaller than one.
The broadening of the kinetic energy histogram is also expected due to the larger contribution of shocks with height, with a high density contrast between shock fronts and expansions.
The average magnetic energy in the upper chromosphere is substantially higher than the kinetic energy as seen in Fig. \ref{fig:energy_avg}.
However, there is still substantial overlap with the wings of the kinetic energy distribution and the magnetic energy distribution. 

\begin{figure*}[htp]
\centering
\includegraphics[width=18cm]{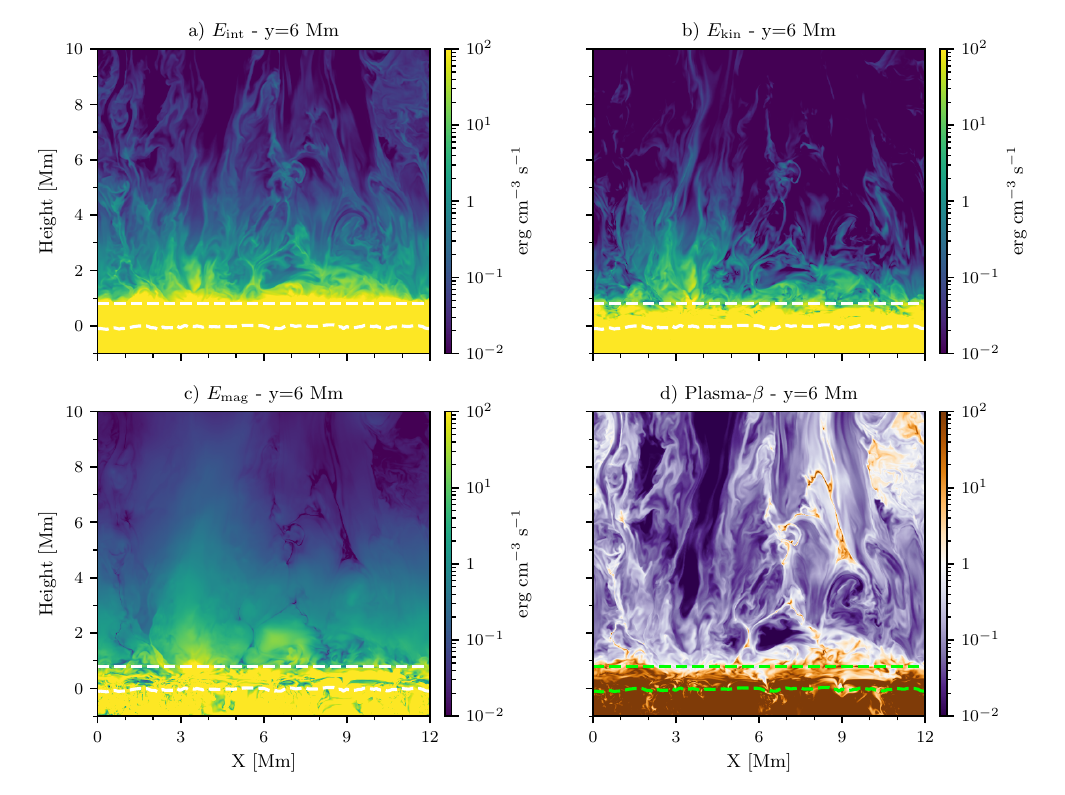}
\caption{The spatial variation of energy density in the xz plane at $y=6~\mathrm{Mm}$, including; panel a) the internal energy b) the kinetic energy, c) the magnetic energy, and d) the plasma beta.
The horizontal dashed lines represent the height of the photosphere ($z=0$) and base of the chromosphere $z=0.8~\mathrm{Mm}$.
An animation of the figure is available online. \href{https://datashare.mpcdf.mpg.de/s/FnnXYXZe99OeYgM}{Animation Available online}, the image represents $t=0$ of the animation.}
\label{fig:Eslice}
\end{figure*}

The magnetic fields in the chromosphere are substantially higher than expected from a potential extrapolation of the photospheric fields.
The distribution of magnetic energy in the upper chromosphere can be better understood by looking at the spatial and temporal variation of the energies through a slice in the simulation domain.
In Fig. \ref{fig:Eslice} we show the internal, kinetic, and magnetic energies as well as the plasma-beta for the $xz$-slice through the atmosphere presented in Fig. \ref{fig:slice_atmosphere}.
The animation of the figure clearly illustrates that shock fronts with high thermal and kinetic energy, where plasma-$\beta \approx 1$, are able to push the magnetic field high into the atmosphere.
This results in a relatively smooth distribution of the magnetic field with height and explains much of the excess field strength relative to a potential field in the upper chromosphere.
Above $\approx 1~\mathrm{Mm}$ the simulation is predominately low-$\beta$, although regions with $\beta \geq 1$ are present in shock-fronts, or in low-field regions of the upper atmosphere.

\subsection{Poynting flux}\label{subsec:poynting_flux}

\begin{figure}[htp]
\centering
\includegraphics[width=8.8cm]{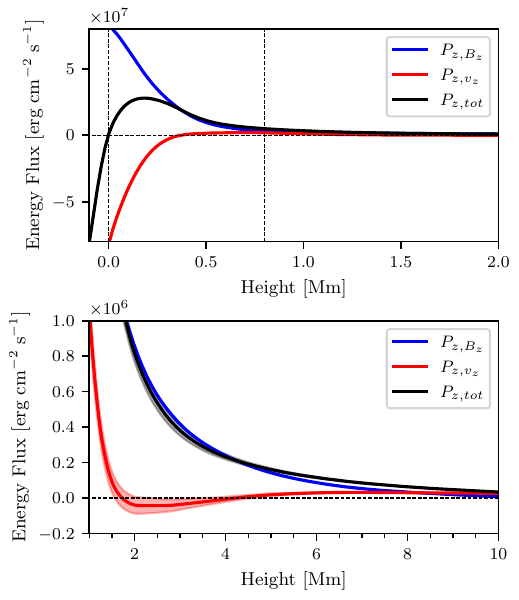}
\caption{Horizontally and temporally averaged values of the  vertical Poynting flux in the solar atmosphere. The top panel shows its stratification in the photosphere and low-chromosphere and the bottom panel displays its behaviour in the chromosphere and above on an expanded vertical scale. The blue line shows the shear (wave) $-B_z ( v_x B_x + v_y B_y)$, the red line the flux advection $v_z B_h$, and the black line the total vertical Poynting flux. The solid lines show the mean value and the shaded area shows two standard deviations.}
\label{fig:pf_havg}
\end{figure}

The magnetic energy at the photosphere represents most of the free energy, making it a strong candidate for heating the chromosphere and corona.
We now investigate the magnetic energy (Poynting) flux into the chromosphere.
Figure \ref{fig:pf_havg} shows the horizontally and temporally averaged vertical component of the Poynting flux $P_z$.
The vertical Poynting flux is further split into two parts, the advection of horizontal field $P_{z,v_z} = \frac{v_z}{4 \pi} (B_y^2 + B_x^2)$, and a term proportional to the vertical field $P_{z,B_z} = -\frac{B_z}{4 \pi} \left(B_x v_x + v_y B_y\right)$.
The latter term represents horizontal motions around a vertical field, including motions with a shear term, braiding, and vortical or wave motions.
Vigorous near-surface convection leads to amplification of magnetic energy near the surface, and re-circulation of this field gives a net negative Poynting flux below the surface.
Above the photosphere, the horizontally and temporally averaged vertical Poynting flux $P_{z}=P_{z,v_z}+P_{z,B_z}$ is positive.
The majority of this Poynting flux is provided through the $P_{z,B_z}$ term.
At the base of the chromosphere, which is the height at which $\left<\tau_{500}\right>=10^{-5}$, the Poynting flux is $\left<P_z\right>_{\mathrm{havg}} = (5.06 \pm 1.08) \times 10^{6} ~\mathrm{erg\;cm^{-2}\;s^{-1}}$.
This is split into the advective term $\left<P_{z,v_z}\right>_{\mathrm{havg}} = (1.86 \pm 1.07) \times 10^{6} ~\mathrm{erg\;cm^{-2}\;s^{-1}}$, and the $\left<P_{z,B_z}\right>_{\mathrm{havg}} = (3.21 \pm 0.12) \times 10^{6} ~\mathrm{erg\;cm^{-2}\;s^{-1}}$.
The $P_{z,v_z}$ term is negative at the photosphere, increasing to a maximum in the low chromosphere before becoming negative through the mid-to-upper chromosphere up to about 4Mm, when it finally turns positive again.

\begin{figure*}[htp]
\centering
\includegraphics[width=16cm]{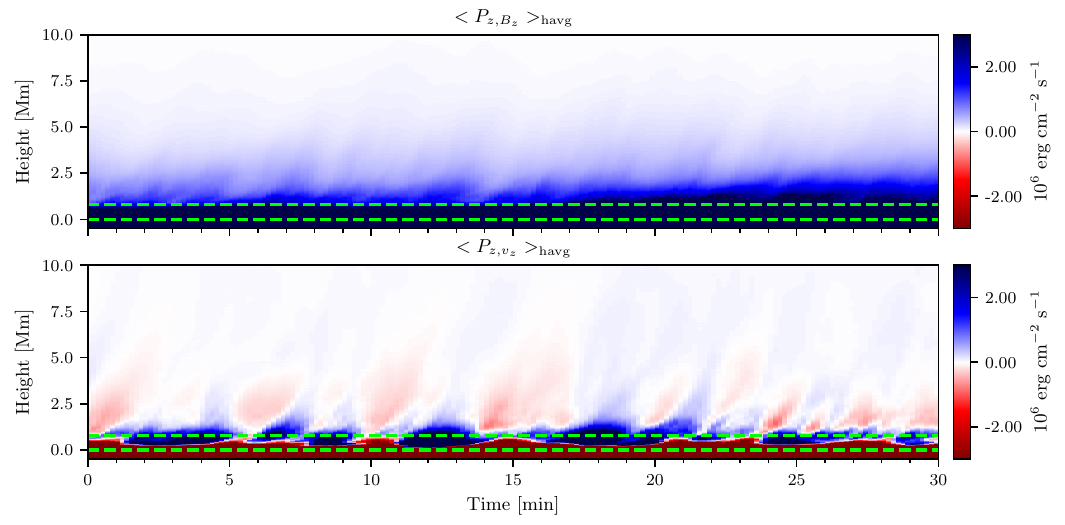}
\caption{Time-distance image of the of the horizontal averages of the a) shear (wave) $-B_z ( v_x B_x + v_y B_y)$, and b) flux advection $v_z B_h$ components of the vertical Poynting flux.
The horizontal dashed lines represent the photosphere $z=0$ and base of the chromosphere $\left<\tau_{500}\right>_{\mathrm{havg}}=10^{-5}$.}
\label{fig:pf_havg_timeseries}
\end{figure*}

The time-dependence of the horizontally averaged Poynting flux is shown in Fig. \ref{fig:pf_havg_timeseries}.
The $P_{z,B_z}$ component of the Poynting flux does not vary strongly with time.
In contrast, the $P_{z,v_z}$ component varies.
Although largely negative in the chromosphere it periodically supplies magnetic energy into the upper chromosphere.
The Poynting flux term representing the vertical advection of horizontal-field $P_{z,v_z}$ varies approximately with the 5-minute turnover time of solar convection.
In a simulation domain with limited size and depth, only a limited range of modes can be excited, and these modes are excited with excess power (see Appendix \ref{app:boxmode}).
The oscillations seen in Fig. \ref{fig:pf_havg_timeseries} are significantly shorter than the fundamental box mode, which has a period of 12 minutes.
Instead, a clear signature is seen with a 5 minute periodicity, which are the acoustic p-modes that propagate throughout the solar interior.
In the upper atmosphere of the simulation the shear (wave)term decreases quicker with height than the advective term.
By $9~\mathrm{Mm}$ the advective term is positive and carries $60\%$ of the total Poynting flux $\left<P_{z,v_z}\right>_{\mathrm{havg}, 9~\mathrm{Mm}} =( 4.71 \pm 1.19 )\times 10^{4} ~\mathrm{erg\;cm^{-2}\;s^{-1}}$.
The reduction in the shear Poynting flux term could be affected by the height of the upper boundary.
A gradient can be seen in the $<P_{z,B_z}>$ component of the Poynting flux with height.
This is caused by a gradual increase of $B_{tot}$ over the timeseries studied.
The average field strength in the chromospheres evolves slowly on a timescale of $\approx 1-2~\mathrm{hours}$, which is not captured completely in the limited time series of this study.
A sharper increase of $B_{tot}$ in the low chromosphere occurs due to a large flux emergence event which occurs around 18 minutes into the simulation.

\begin{figure*}[htp]
\centering
\includegraphics[width=16cm]{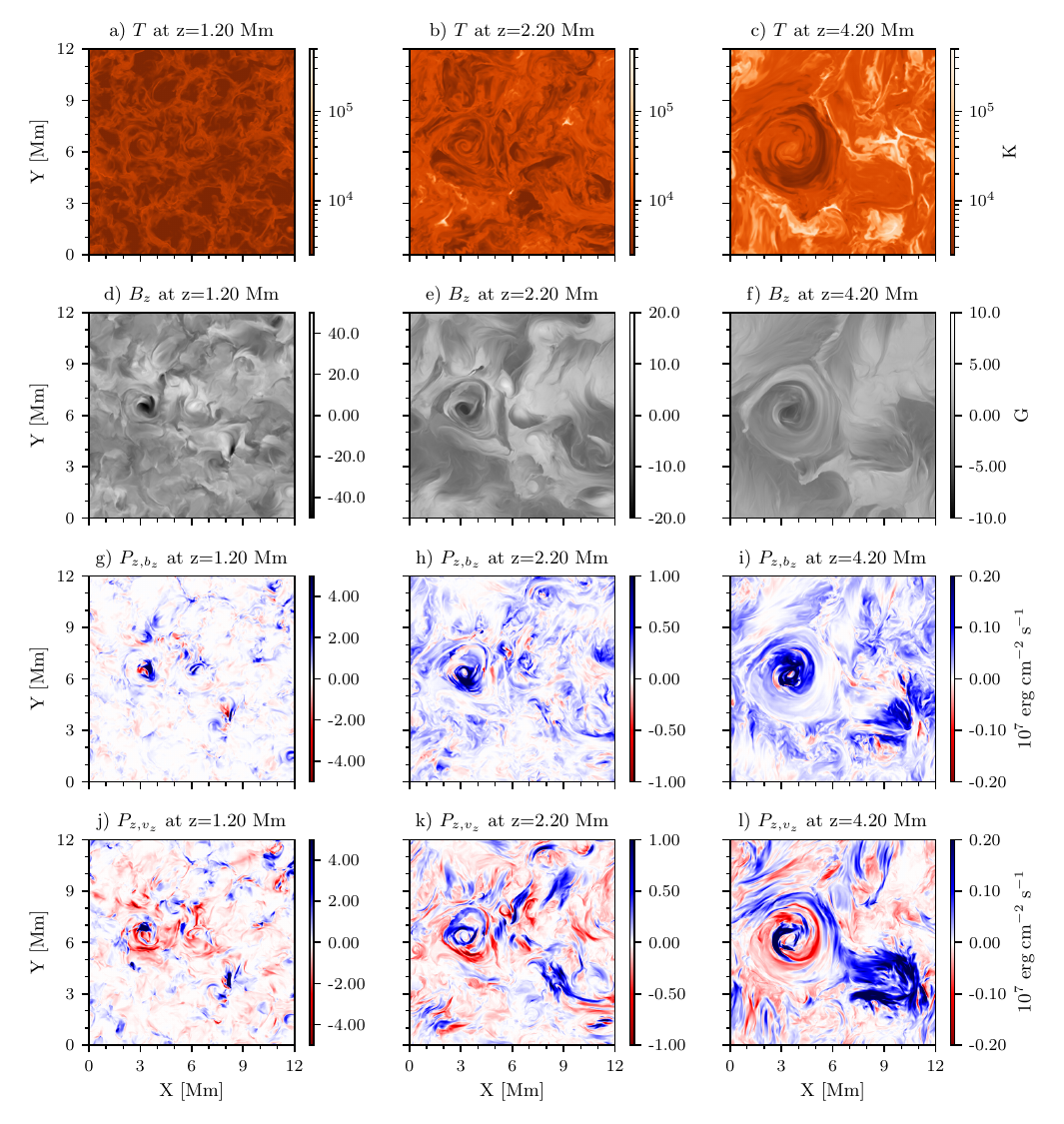}
\caption{Horizontal cuts in the chromosphere, left-to-right, at heights of $z=1.2,\;2.2~\mathrm{and}~4.2~\mathrm{Mm}$ above $z=0~\mathrm{Mm}$. The panels (top-to-bottom) include the temperature, vertical magnetic field $B_z$, and two components of the vertical Poynting flux $P_{z,B_z} = -B_z ( v_x B_x + v_y B_y)$ , and $P_{z,v_z} = B_h$. The temperature panels are limited to compare the structures at different layers in the chromosphere. In panel c) approximately 8 pixels in a million are above $1~\mathrm{MK}$. The field strength and Poynting flux panels are limited to highlight the chromospheric structures. \href{https://datashare.mpcdf.mpg.de/s/UfYFV6H2FHsBNxp}{Animation Available online}, the image represents $t=0$ of the animation.}
\label{fig:T_B_pf}
\end{figure*}

Finally, we aim to understand the spatial distribution of the Poynting flux in the internetwork chromosphere.
Horizontal slices through the chromosphere are shown in Fig. \ref{fig:T_B_pf} for three heights, at $1.2~\mathrm{Mm}$, $2.2~\mathrm{Mm}$, and $4.2~\mathrm{Mm}$ above $z=0~\mathrm{Mm}$.
An animation of the figure is available.
The top row shows the temperature at these different heights. In the lower chromosphere (panel a), the temperature shows a shock interference pattern, with small regions of sustained high temperature present above strong field concentrations.
By $2.2~\mathrm{Mm}$ the plasma-$\beta$ is low and the dynamics begin to follow the magnetic field, becoming highly structured perpendicular to the magnetic field.
Most of the plasma is around $20 ~\mathrm{kK}$, with large pockets of low-chromospheric temperatures, and small regions reaching transition region temperatures.
The plasma is highly structured in the horizontal direction with sharp variations from swirling motions and shock buffeting.
In the highest slice most of the plasma remains at upper chromospheric temperatures of around $20~\mathrm{kK}$, with some regions being heated to reach coronal temperatures, nearly $1~\mathrm{MK}$.
A large swirling event is seen in panel c), at around $x=4~\mathrm{Mm},y=6~\mathrm{Mm}$, with significant low chromospheric material reaching high in the atmosphere.
It is also seen at 2.2 Mm, although less clearly.
The swirl is easier to detect at all heights in other physical parameters, such as $B_z$ or Poynting flux.

The second row shows the vertical magnetic field strength.
In the low chromosphere a number of small, sub-granular loops are seen, with field strengths above $50~\mathrm{G}$.
These loops can be identified in the horizontal slices by a smooth transition of polarity along the structure.
At $2~\mathrm{Mm}$ the field is dominated by loops on the granular scale, and the field has dropped to a maximum value of around $10~\mathrm{G}$.
At $4~\mathrm{Mm}$ the field shows a large scale flux-imbalance of around $4~\mathrm{G}$ over horizontal scales of a few granules ($4-8~\mathrm{Mm}$).
Even in the upper chromosphere significant small scale structure is seen in the magnetic field, down to $\approx 100~\mathrm{km}$ with opposite polarity loops frequently interacting. 

The bottom two rows of Fig. \ref{fig:T_B_pf} show the $P_{z,v_z}$ and $P_{z,B_z}$ components of the Poynting flux.
The animation of Fig. \ref{fig:T_B_pf} reveals the dynamics which lead to strong Poynting flux in the chromosphere.
A swirl at $x=4~\mathrm{Mm}, y=7~\mathrm{Mm}$ expands into the upper chromosphere and carries significant Poynting flux upwards.
The swirl is present at the beginning of the simulation and remains visible for approximately $20~\mathrm{min}$ before starting to decay.
A number of smaller-scale swirls can be seen in the simulation, but none has as large an impact on the upper chromosphere. 
At $x=8~\mathrm{Mm}, y=4~\mathrm{Mm}$ the interaction of a number of small loops carries significant Poynting flux into the upper chromosphere. 
Most of the other regions of enhanced Poynting flux are more transient, lasting only minutes. 

\section{Discussion} \label{sec:discussion}

Whereas previous work included an imposed vertical field, which dominates the field geometry of at least the higher layers of the chromosphere, in the present work, we produce a model of the quiet sun internetwork generated through a SSD mechanism.
The SSD, acting in the upper convection zone of the simulation, generates magnetic fields consistent with photospheric observations of the internetwork field, as has been showed in a number of papers, e.g. \citep{danilovic_2010_MURaMHinode_QS_LD,danilovic_2016_SSDHinode,lites_2008_hinode_QSIN,lites_2011_hinodeSSD,JTB_2004_hanlefield,dPA_2018_Hanle_QS,zeuner_2024_hanle}
We study the strength of fields in the chromosphere, and investigate their contribution to the chromospheric energy balance. 
We demonstrate that a zero net-flux SSD simulation is capable of generating magnetic fields that reach high into the chromosphere.
The small simulation domain ($12\times12~\mathrm{Mm}$) restricts the formation of larger-scale magnetic loops reaching high into the atmosphere and likely dominating the magnetic structure there. 
Even so, the magnetic field in the upper chromosphere is strong enough that the magnetic energy is more than an order of magnitude higher than the kinetic energy, and around a factor of 2 higher than the thermal energy.

We find that the magnetic energy  falls off far more slowly with height than expected from a potential field extrapolation.
A mechanism for the enhanced field strengths in the chromosphere is shown to be the interaction between shocks and the magnetic field.
The shock-fronts are seen to push the turbulent internetwork fields higher into the atmosphere (see for example the animation of Fig. \ref{fig:Eslice}). 
Although the shocks in the chromosphere act to locally enhance the magnetic field, this does not constitute a dynamo process.
It is possible for local enhancement of the field to occur without a dynamo process.
An additional source of energy is contained in the braiding of the field, caused by small-scale vortical motions, which is ubiquitous in simulations of the solar atmosphere.

These changes will cause extrapolations of quiet inter-network regions to significantly underestimate the magnetic field strength in the chromosphere.
This will further impact the inferred pressure balance, density, and free magnetic energy present in the chromosphere.
Finally the height of the network canopy will be increased due to the higher pressure in the internetwork regions.

At the photosphere, the behaviour of quiet sun internetwork fields have been well studied in both observations \citep{buehler_2013_hinodeSSD,gosic_2014_IN_1,gosic_2016_IN2,lagg_2016_photospheric_QS,smitha_2017_QS_FE} and through numerical simulations \citep{danilovic_2010_MURaMHinode_QS_LD,rempel_2014_numerical,bhatia_2022_SSD,bhatia_2023_SSD2}.
Due to the difficulties in measuring chromospheric magnetic fields, and the complexity of detailed chromospheric simulations, the extent to which these fields impact the chromosphere has remained poorly understood.
\citet{khomenko_2018_SSD} performed a simulation including the low chromosphere, with an upper boundary at $1.4~\mathrm{Mm}$.
These simulations study the propagation of magnetic energy including non-ideal effects such as ambipolar diffusion and the Hall effect.
They find a $90\%$ reduction of Poynting flux before the upper boundary of the simulation.
However the upper boundary is too low, so that it is expected to significantly affect the field topology in the chromosphere.
Previous numerical studies of the internetwork chromosphere, including high resolution and an upper boundary above the transition region, are limited to the model of \citet{martinez_sykora_2019_chromo_field} (MS19).

Observational inference of weak quiet sun fields, in the range of $1~\mathrm{to}~100~\mathrm{G}$ in the chromosphere, typically requires Hanle depolarisation measurements \citep{stenflo_1982_solar_hanle,JTB_2022_Hanle_review}. \citet{Faurobertscholl_1992_chromo_hanle} investigated depolarisation of the \ion{Ca}{i}$\;4227~\AA$ line, which forms in the mid chromosphere, roughly $700-1200~\mathrm{km}$ above the solar surface in a 1D plane parallel solar atmosphere model.
Assuming a height-dependent formation due to the chromospheric canopy they found a $20-100~\mathrm{G}$ field strength.
The observations of \citet{bianda_1998_chromo_hanle}, of the same line, measure an average of $5-15~\mathrm{G}$ value over the formation height of the line.
\citet{lagg_2009_10830_hanle} investigated Hanle depolarisation of the \ion{He}{i}$\;10830~\AA$ line and found a field strength of $20-50~\mathrm{G}$ in a weak field region.
These results are consistent with those in our simulation, with an average field strength of $\approx 15~\mathrm{to}~30~\mathrm{G}$ found over the formation region $700-1200~\mathrm{km}$.

A previous study of the quiet sun chromospheric field by MS19 found similar properties of the magnetic field in the low atmosphere.
They find the magnetic field is strongly horizontal in the low-chromosphere.
Their work included a $2.5~\mathrm{G}$ vertical guide field, leading to the field becoming vertical in the upper chromosphere.
The internetwork SSD simulation presented in this work does not include any imposed field, and the limited domain size prevents the formation of large-scale network-fields.
This causes the chromospheric field to remain highly inclined through the bulk of the chromosphere, in contrast to MS19.
We see significantly stronger magnetic energy in the chromosphere, around a factor of 2 higher than MS19.
This is likely caused by the larger, deeper convection zone in combination with the reduced diffusivity of the slope-limited numerical diffusion scheme, leading to a stronger dynamo-generated magnetic field in the photosphere. In the internetwork SSD simulation there is a close equipartition of kinetic and magnetic energy in the low chromosphere and magnetic energy begins to dominate in the mid-chromosphere.
The kinetic energies in the two simulations are of a similar magnitude at $1~\mathrm{Mm}$, which is a factor of $\approx 4$ larger than the magnetic energy in MS19.
By the upper chromosphere the $2.5~\mathrm{G}$ imposed field begins to dominate and the two simulations can no longer be directly compared.

The measured Poynting fluxes in this simulation are consistent with those presented in \citet{tilipman_2023_photospheric_PF}.
They present simulations, performed with the MURaM code, include the upper convection zone and low chromosphere of a SSD-driven quiet sun simulation.
Their treatment of radiation and hydrogen ionisation in the chromosphere is limited to LTE.
They observe a turnover in the Poynting flux just below the photosphere and a Poynting flux peak of $2.28\times10^7~\mathrm{erg\;cm^{-2}\;s^{-1}}$ at $128~\mathrm{km}$.
Comparing MURaM simulations to observational data they find that interpretation of the observations (see also \citet{silva_2022_PFobs}) is complicated by the measurement of the field on an optical depth surface. Additionally, it is difficult to measure both the vertical and horizontal field components that are required to determine the shear (wave) component ($P_{z,B_z}$). We significantly extend the study of the photospheric Poynting flux by investigating the propagation of energy into the solar chromosphere, where we treat non-equilibrium hydrogen ionisation and non-LTE radiative transfer.

In order to understand the energy flux propagating from the turbulent convection zone into the atmosphere we take spatial and temporal averages at the base of the chromosphere.
The spatially and temporally averaged vertical Poynting flux into the chromosphere is $\left<P_z\right>=5.01\times 10^6 ~\mathrm{erg\;cm^{-2}\;s^{-1}}$, approximately 25\% higher than the canonical energy flux required to counteract radiative losses in the quiet sun chromosphere ($4.3\times 10^6~\mathrm{erg\;cm^{-2}\;s^{-1}}$) and corona ($3\times 10^{5}~\mathrm{erg\;cm^{-2}\;s^{-1}}$) \citep{Withbroe_1977_chromoheating}.
This is consistent with large simulations, such as \citet{rempel_2017_extension,chen_2021_campfires} which supply an average vertical Poynting flux of approximately $5 \times 10^5~\mathrm{erg\;cm^{-2}\;s^{-1}}$ to the corona.
Therefore, the action of fields self-consistently generated by an SSD, convective turbulence easily provides a large enough energy flux into the chromosphere.
We have demonstrated that this is the case even for zero-net magnetic flux.

In order to determine the complete energy balance of the modelled chromosphere we additionally calculate the vertical kinetic energy flux, $F_{z,E_{\mathrm{kin}}}=1/2\rho v_z \left(v_x^2+v_y^2+v_z^2\right)$ and enthalpy flux $F_{z,H}= v_z \left(E_{\mathrm{int}}+p\right)$.
The averaged kinetic energy flux into the chromosphere, $<F_{z,E_{\mathrm{kin}}}>_{\mathrm{avg}}= -4.8\times 10^5~\mathrm{erg\;cm^{-2}\;s^{-1}}$, is an order of magnitude smaller than the Poynting flux and negative.
The averaged vertical enthalpy flux at the base of the chromosphere is high $\left<F_{z,H}\right>_{\mathrm{havg}}=3.63\times 10^6~\mathrm{erg\;cm^{-2}\;s^{-1}}$.
This value is highly sensitive to the height at which the energy flux is measured.
It is negative below $450~\mathrm{km}$, with a peak at $600~\mathrm{km}$, falling an order of magnitude by 1 Mm.
This strong peak is presumably related to the height of the canopy in the low chromosphere \citep[e.g.][]{solanki_1990_magneticchromosphere}, the sharp decrease in Poynting flux in the upper photosphere leads to an upwards enthalpy flux in the low chromosphere.
To complete the energy balance, we include radiative losses in the chromosphere $\sum_{z=0.8~\mathrm{Mm}}^{10~\mathrm{Mm}} \left<Q_{rad}\right>_{\mathrm{havg}} dz=-8.44\times 10^6~\mathrm{erg\;cm^{-2}\;s^{-1}}$.
Here $dz$ is the vertical resolution and $Q_{rad}$ is the total radiative losses/heating rate from the multigroup scheme, chromospheric lines, optically thin losses, and EUV back heating.
These losses are nearly double those expected from the values of \citet{Withbroe_1977_chromoheating}.
Possible causes of these differences could be the difficulty in defining the `base of the chromosphere' in a dynamic, 3D model, long term variations of the chromosphere, or the lack of a canopy and network in the simulation.

Attributing chromospheric heating to a particular energy flux at the base of the chromosphere is difficult, and would require a more detailed study than presented in this work.
The RMS values of both components of the Poynting flux and the kinetic energy flux are of a similar magnitude, but the mean values differ significantly. 
The shear ($P_{z,B_z}$) component of the Poynting flux and the enthalpy flux are the dominant contributors of energy into the chromosphere.
An approach such as the Lagrangian 'corks' described by \citet{leenaarts_2018_corks} could help to understand the mass transfer between the photosphere, chromosphere and corona.
However, tracing the transport of energy carried by Alfv\'enic waves is complicated by the complex topology and frequent restructuring of the chromospheric magnetic fields.
The dissipation of these energies, and the region of the chromosphere in which they are thermalised will be dependent on the resolution and numerical diffusion scheme. 
Additional sources of energy dissipation, for example through ion-neutral or multi-fluid effects \citep{khomenko_2017_review_nonideal,martinezsykora_2012_PI_1,martinezsykora_2020_NE_PI} will change the dissipation of kinetic and magnetic energy.

We do not include a generalised Ohm's law to model the effect of ambipolar diffusion.
\citet{khomenko_2018_SSD} perform a LTE simulation of SSD generated fields, including a generalised Ohm's law, with an upper boundary in the low chromosphere ($1.4~\mathrm{Mm}$).
In their simulation the inclusion of ambipolar diffusion leads to a significant dissipation of Poynting flux in the solar chromosphere.
The inclusion of non-equilibrium ionisation could significantly change these results as hydrogen is the major electron donor in the chromosphere.
The study of MS19 was extended to include NLTE effects by \citet{martinez_sykora_2023_IN_NE_GOHM}, where the inclusion of NE ionisation and a generalised Ohm's law leads to increased temperatures in the upper chromosphere. 
However, they do not include results from a simulation including NE ionisation without a generalised Ohm's law, to allow a direct comparison to this study.
In the presented simulation the energy into the chromosphere is a conserved quantity, with the largest diffusivity dissipating the most energy.
Only the details of the distribution of energy dissipation between the different terms will change as resolution is increased, or if non-ideal effects are included.
The results presented in this paper are therefore expected to be robust.

The extent of the numerical domain is insufficient in depth and width to self-consistently model supergranulation.
The horizontal motions at scales of supergranulation ($\approx 20\Mm$) advect field to edges of the supergranules.
This produces a net flux imbalance on these scales, and network loops will reach higher and likely contribute to heating the plasma in higher layers to coronal temperatures. 
As seen in Fig \ref{fig:energy_avg}, the scale height of the vertical magnetic field in the upper atmosphere is roughly equal to $k=2\pi/6 ~\mathrm{Mm}$.
In a larger SSD simulation, such as \citep{chen_2021_campfires,chen_2022_dopplershifts}, the magnetic fields emerging in the internetwork interact with these large scale fields. 
The horizontally averaged RMS field strength is $\approx 3~\mathrm{G}$ at $4\Mm$ above the photosphere (see Fig. \ref{fig:magnetic_field_avg}).
A typical super-granular flux imbalance is approximately $\pm 2.8 ~\mathrm{G}$ \citep{smitha_2017_QS_FE}.
In the upper-chromosphere, the dynamics seen in the current simulation will become dominated by the magnetic canopy provided by the magnetic network. 
The emerging internetwork fields, which currently propel material to the top of the simulation domain, will interact with these overlying fields. 
The interaction between the highly turbulent internetwork fields and canopy will drive reconnection events, further heating the chromosphere and corona.

\section{Conclusion} \label{sec:conclusion}

In this work we demonstrate that even in the quietest regions of the sun the magnetic field is strong enough to dominate the energy balance above the mid-chromosphere.
The kinetic and magnetic energies are in equipartition throughout the low chromosphere,  while in the higher layers the magnetic energy density is an order of magnitude higher. In the higher layers the magnetic energy is also a factor of 2 larger than the thermal energy density.
This equipartition is caused by the advection and compression of field by shocks in the lower atmosphere, and the anisotropy of the field and hot plasma in the upper atmosphere.
The resulting field provides a Poynting flux strong enough to heat the chromosphere.
Additionally the magnetic canopy in the upper photosphere produces a significant enthalpy flux into the chromosphere.

However the resulting simulation is unable to form a steady million degree corona. Only small, fluctuating pockets of million-degree gas are formed. 
This is likely due to the too low magnetic energy at greater heights in our simulation.
Within the limited spatial domain of the simulation, a magnetic network is not formed.
Therefore, no strong magnetic loops which reach high into the atmosphere are present in the simulation.
Larger loops will provide a mechanism to allow a larger Poynting flux to reach higher into the atmosphere.

\begin{acknowledgements}

D.P. would like to thank T. Bhatia for discussions on the SSD. This project has received funding from the European Research Council (ERC) under the European Union’s Horizon 2020 research and innovation programme (grant agreement No. 101097844 ). This work was supported by the Deutsches Zentrum f{\"u}r Luft und Raumfahrt (DLR; German Aerospace Center) by grant DLR-FKZ 50OU2201. We gratefully acknowledge the computational resources provided by the Raven supercomputer of the Max Planck Computing and Data Facility (MPCDF) in Garching, Germany. D.P. would like to thank A. Irwin (Free-EoS) and V. Witzke (MPS-ATLAS). This material is based upon work supported by the NSF National Center for Atmospheric Research, which is a major facility sponsored by the U.S. National Science Foundation under Cooperative Agreement No. 1852977. This project has received funding from the Swedish Research Council (2021-05613) and the Swedish National Space Agency (2021-00116).

\end{acknowledgements}

\bibliography{references}

\begin{appendix}

\section{Simulation setup}\label{app:simulation}

In this section we provide a summary of the numerical setup used in the simulation. Table \ref{tab:sim_setups} includes information on the numerical scheme, boundary conditions and physics modules included. In Table \ref{tab:governing_parameters}  we provide the governing parameters of the run. These include the simulation size and resolution, and the net magnetic flux. The type of star which will be simulated is set by the inflowing entropy and pressure at the lower boundary, the gravity and elemental abundances.

\begin{table}[htp]
    \caption{Simulation setups}
    \centering
    \begin{tabular}{|c | c |}
    \hline
    Simulation setup & description \\
    \hline
        Spatial derivatives & 4th order central differences \\ 
    \hline
        Time-integration & 4-stage explicit update \tablefootmark{a}  \\
	    \hline
        Diffusive scheme & Slope-limited numerical diffusion \tablefootmark{b}  \\
	    \hline
        Upper BC & Open to inflows, closed to outflows \\
         & Potential magnetic field \\
	    \hline
        Lower BC & Open boundary condition  \\
                                 & Symmetric for magnetic field \\
	    \hline
        Horizontal BC & Periodic \\
	    \hline
        EoS & LTE in solar interior and photosphere\\
            & NE in chromosphere and corona\\
	    \hline
        Radiation transfer &  Scattering multi-group scheme \\
         &  Tabulated chromospheric losses \\
         &  Tabulated optical thin losses \\
         &  3D EUV back-heating \\
	    \hline
        \end{tabular}
    \label{tab:sim_setups}
    \tablefoot{
    \tablefoottext{a}{\citet{jameson_2017_timeupdate}}
    \tablefoottext{b}{\citet{rempel_2014_numerical,rempel_2017_extension}}
    }
\end{table}

\begin{table}[htp]
    \caption{Governing parameters of the simulation.}
    \centering
    \begin{tabular}{|c | c |}
    \hline
    Parameter & Value  \\
    \hline
        nx & 512  \\ 
        ny & 512  \\
        nz & 900  \\ 
    \hline
        dx & 23.4375 km\\ 
        dy & 23.4375 km\\
        dz & 20 km\\ 
     \hline
        $<B_z>_{\mathrm{havg}}$ & $0~\mathrm{G}$   \\       
    \hline
        $s_{bc}$ & $1.41171\times10^9$ erg$\;$cm$^{-3}\;$K$^{-1}$\\
	    \hline
        $p_{bc}$ & $1.77668\times10^9$ g$\;$cm$^{-1}\;$s$^{-2}$  \\
	    \hline
        $g$ & $2.74\times10^4$ cm$\;$s$^{-2}$ \\
	    \hline
        Abundances & Asplund 2009 \tablefootmark{a}\\
	    \hline
        \end{tabular}
    \label{tab:governing_parameters}
    \tablefoot{
    \tablefoottext{a}{\citet{asplund_2009_abu}}
    }
\end{table}

\section{Simulation box mode}\label{app:boxmode}
\begin{figure}[htp]
\centering
\includegraphics[width=8.8cm]{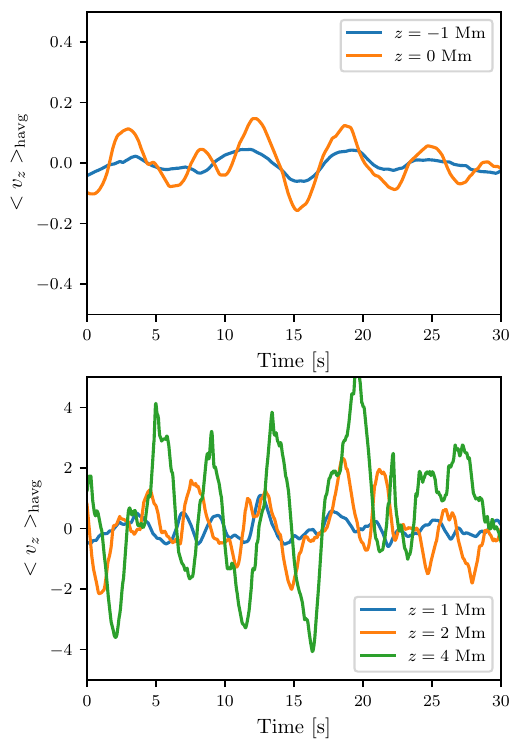}
\caption{Variation of the horizontally averaged vertical velocity in the simulation. The figure shows different heights in the interior and near-photosphere (top panel), and the chromosphere (bottom panel) .}
\label{fig:boxmode}
\end{figure}

Turbulent near-surface convection excites modes with a range of frequencies and wavenumbers. In a simulation of limited width and depth, with a partially absorbing lower boundary, the power is spread over a limited set of modes. In particular, a global box mode can be excited with excess power. To visualise this effect, we plot in Figure \ref{fig:boxmode} the horizontally averaged vertical velocity in the simulation, see also Fig. 8 of \citet{carlsson_2016_public}. At photospheric heights a periodicity of approximately 5-6 minutes is seen. The period of the fundamental node can be approximated by calculating the travel time of an acoustic wave between the photosphere to the lower boundary of the simulation $T=2 \int_{z=-6.8}^{0} \frac{dz}{C_s}$, where $C_s$ is the horizontally averaged sound speed. This gives a value of 12.8 minutes, which is significantly longer than the dominant oscillation observed in Fig. \ref{fig:boxmode}.

\end{appendix}
\end{document}